\documentclass[prb,twocolumn,showpacs,showkeys,superscriptaddress]{revtex4-2}

\usepackage{amsmath,amssymb,amsfonts,dsfont,mathrsfs,amsthm}
\usepackage{graphicx}
\usepackage{centernot}
\usepackage{hyperref}
\usepackage[utf8]{inputenc}
\usepackage{xcolor}
\usepackage{comment}
\hypersetup{linktocpage,colorlinks=true,urlcolor=blue!80!red,linkcolor=blue,citecolor=red}
\usepackage{feynmf}
\usepackage{siunitx}
\usepackage{array}
\usepackage{ulem}
\usepackage{tikz}
\usepackage{braket}

\begin{document}

\title{High - $T_{c}$ Berezinskii-Kosterlitz-Thouless transition in 2D superconducting systems with coupled deep and quasi-flat electronic bands  with van Hove singularities}
\author{Sathish Kumar Paramasivam}
\address{School of Science and Technology, Physics Division, University of Camerino, 62032 Camerino (MC), Italy.}
\address{Department of Physics, University of Antwerp, Groenenborgerlaan 171, 2020 Antwerp, Belgium.}
\author{Shakhil Ponnarassery Gangadharan}
\address{School of Science and Technology, Physics Division, University of Camerino, 62032 Camerino (MC), Italy.}
\address{Department of Physics, University of Antwerp, Groenenborgerlaan 171, 2020 Antwerp, Belgium.}
\author{Milorad V. Milo\v{s}evi{\'c}}
\email{milorad.milosevic@uantwerpen.be}
\address{Department of Physics, University of Antwerp, Groenenborgerlaan 171, 2020 Antwerp, Belgium.}
\affiliation{Instituto de Física, Universidade Federal de Mato Grosso, Cuiabá, Mato Grosso 78060-900, Brazil.}
\author{Andrea Perali}
\email{andrea.perali@unicam.it}
\address{School of Pharmacy, Physics Unit, University of Camerino, 62032 Camerino (MC), Italy.}

\begin{abstract}
In the pursuit of higher critical temperature of superconductivity, quasi-flat electronic bands and van Hove singularities in two dimensions (2D) have emerged as a potential approach to enhance Cooper pairing on the basis of mean-field expectations. However, these special electronic features suppress the superfluid stiffness and, hence, the Berezinskii-Kosterlitz-Thouless (BKT) transition in 2D superconducting systems, leading to the emergence of a significant pseudogap regime due to superconducting fluctuations. In the strong-coupling regime, one finds that superfluid stiffness is inversely proportional to the superconducting gap, which is the predominant factor contributing to the strong suppression of superfluid stiffness. Here we reveal that the aforementioned limitation is avoided in a 2D superconducting electronic system with a quasi-flat electronic band with a strong pairing strength coupled to a deep band with weak electronic pairing strength. Owing to the multiband effects, we demonstrate a screening-like mechanism that circumvents the suppression of the superfluid stiffness. We report the optimal conditions for achieving a large enhancement of the BKT transition temperature and a substantial shrinking of the pseudogap regime by tuning the intraband couplings and the pair-exchange coupling between the two band-condensates.
\end{abstract}
\maketitle
\section{Introduction}
In 2D superconductors, the Berezinskii-Kosterlitz-Thouless (BKT) transition temperature $T_{BKT}$ \cite{bkt,bkt1} determines the phase transition from a normal state to a superconducting state with a quasi-long range order. Below $T_{BKT}$, the system exhibits superconducting behavior, whereas above $T_{BKT}$, superconductivity is lost because of dissipation induced by vortex-antivortex unbindings \cite{petter}. The BKT transition temperature can be evaluated from the superfluid stiffness ($\rho_{s}$) \cite{kt}. The superfluid stiffness of conventional 2D superconductors with a single electronic band has an inverse relationship with the effective mass $m^{*}$ of the band \cite{pratap}. This suggests that the superfluid stiffness is very small when the band is quasi-flat, highlighting the crucial role of band geometry in determining the BKT transition temperature of 2D superconducting systems. In this work, our focus will be on circumventing this problem within the framework of multiband superconductivity. Multiband superconductors often possess novel properties that are not present in their single-band counterparts \cite{2band,tanaka,s.z.lin}. The concept of superfluid stiffness in multiband superconductors can become more complex, as different bands may have different energy gaps, effective masses, and presence of saddle points in the electronic dispersion, exhibiting different levels of superfluid stiffness. The pair-exchange couplings of the band condensates can also affect the total superfluid stiffness. As a result, understanding the superfluid stiffness in multiband superconductors is important for characterizing their properties and predicting their behavior to optimize configurations for high-temperature 2D superconductivity. In multiband superconductors comprising two distinct electronic bands, there is potential for a significant enhancement of the superconducting energy gaps and a higher critical temperature ($T_{c}$) at a mean-field level \cite{giant_shanenko, y.chen, D.Innocenti,m.v.m,c.yue}. It has been recently shown that the existence of pair-exchange coupling between distinct bands can minimize the superconducting fluctuations through multiband screening processes \citep[][]{screening,multi,suppression}. A significant motivation to study the BKT transition in a two-band system having van Hove singlularities together with deeply dispersive bands stems from the superconducting properties exhibited by monolayer FeSe superconductors. Monolayer FeSe on strontium-titanate (STO) is considered one of the most intriguing iron-based superconductors, primarily due to its remarkably high critical temperature, exceeding 100 K \cite{wang,He,Ge,hoffman}. In contrast, the bulk form of FeSe demonstrates a significantly lower superconducting critical temperature of 8 K \cite{pnas}. Additionally, it is anticipated that bulk FeSe possesses a van Hove singularity at the M point \cite{K.nakayama, coldea}. Conversely, monolayer FeSe, characterized as a two-band superconductor, presents a low-energy electronic structure featuring both an electronic band crossing the Fermi level and an incipient band \cite{Hirschfeld, incipient} that can cooperate in the pairing process. The growth of FeSe monolayer on distinct substrates results in a quasi-flat and deep band formation within the electronic band structure \cite{zhao, xue, m.l.cohen}. Recently, experimental observation of the BKT transition has been reported in this system. 1-unit-cell thick FeSe films exhibit a BKT transition at 23.1~K \cite{zhang}, while the transition temperature for thin flakes of FeSe was measured to be 2.9~K for a sample with a thickness of 14~nm and 6.67~K for a sample with a sizeable thickness of 100~nm \cite{Farrar}, and the values of $T_{BKT}$ have been observed for different thicknesses of the FeSe film \cite{Schneider}. In all of the aforementioned FeSe samples, the analysis of voltage-current ($V(I)$) characteristics unveils a distinct signature of a BKT transition, characterized by a $V\propto I^{\alpha}$ power law dependence. Notably, the exponent $\alpha$ tends towards 3 at $T_{BKT}$ \cite{caprara, sharma}.\\

An additional motivation for this work arises from the effective two-gap/patch model discussed in Refs. \onlinecite{two_patch_1, two_patch_2} in connection with the phenomenology of cuprate superconductors. The effective two-gap/patch model encompasses two distinct wave-vector regions of the band dispersion characterized by different intraband and pair exchange interactions. One of the bands demonstrates a large Fermi velocity and a weak attraction, resulting in the formation of overlapped Cooper pairs with minimal superconducting fluctuations. Conversely, the other band showcases a small Fermi velocity and a strong attraction, leading to formation of tightly bound pairs with strong fluctuations. Most relevantly for this work, the patches of the Fermi surface around the M points of the Brillouin zone (BZ) of cuprates contain a branch of the electronic dispersion characterized by a saddle point, which in 2D is responsible for the van Hove singularity in the density of states. In the two-gap/patch effective model for cuprates, there is a dynamic interplay that emerges between the anisotropic Fermi surface and the wave vector-dependent pairing interaction induced by the charge-density wave (CDW) fluctuations, leading to very different pairing strengths in different arcs of the Fermi surface. This is the electronic structure and pairing configuration that we intend to investigate in this work. The recent investigations into multigap superconductivity in lithium-intercalated bilayer Mo$_{2}$C, conducted through first-principles calculations, show that such intercalation, accompanied by a 3\% tensile strain, results in a notable elevation of the critical temperature ($T_{c}$) to 24~K \cite{mo2c}. This enhancement of superconductivity induced by the strain primarily stems from the downward shift of an energy band with a flat dispersion to energy levels near the Fermi level. A pivotal observation is the coexistence of a flat and a deep dispersive electronic band around the $\Gamma$ point of the Brillouin zone. Furthermore, in multiorbital model of  alkali-doped fullerides (A$_{3}$C$_{60}$), the \textit{ab initio} calculations of the band structure reveal the coexistence of a quasi-flat and a deep dispersive band in the BZ, as shown in the supplementary materials of Ref. \cite{A3C60}. Therefore, the aforementioned 2D multiband superconductors bear qualitative correspondence with the model we have adopted in our study.

To incorporate the essential characteristics of the electronic band structure of a FeSe monolayer into our model, we utilize a two-band tight-binding approach that effectively represents the fundamental aspects of the FeSe monolayer's electronic structure \cite{werner,ashwin,raghu,rafael}. Specifically, we consider a two-band electronic 2D system with a quasi-flat band and a deep band, together with multichannel pairing interactions able to induce strong-coupling multi-gap superconductivity. Remarkably, the band features and density of states emerging from the two-band tight-binding model closely resemble those of the minimal two-band model discussed in the Refs. \cite{werner,ashwin,raghu,rafael}. This correspondence suggests that our adopted two-band tight-binding model is likely to capture the distinctive characteristics of the FeSe monolayer, which are key for evaluating the superconducting properties of the system.  In this work, our objective was to propose a broadly relevant  two-band system characterized by the coexistence of quasi-flat and deep bands in the same 2D superconducting system, in order to produce a condensate with a mixture of BCS-like and BEC-like partial condensates, offering the optimal conditions to stabilize BKT transitions at high critical temperature. Particular emphasis is placed on the physically relevant configuration involving a quasi-flat band (with intra-band pairing ranging from weak to strong) coupled with a deep band (with weak intra-band pairing). We place focus on the effect of the pairing interactions and the van Hove singularity of the quasi-flat band on the BKT transition temperature of the two-band 2D superconductors. Weak coupling strengths in the deep band and even a small pair-exchange coupling are able to suppress the pseudogap regime and stabilize high values of the BKT temperature of the two-band 2D system. In this manner, we demonstrate a unique screening-like mechanism acting on the BKT transition. In the framework of the examined two-band electronic system, we have explicitly accounted for both intraband couplings, wherein Cooper pairs undergo creation and annihilation within the same band, and pair-exchange interband coupling, where Cooper pairs are created in one band and annihilated in the other. A formal transformation of the multi-orbital BCS mean-field Hamiltonian from an orbital basis representation to a band basis representation is undertaken, leading to pair-exchange and cross-pair interband interactions in the band representation. The same mechanism is at work when a real-space Hamiltonian with local interactions is diagonalized in a band representation, with the emergence of pairing amplitudes in all channels, as intraband and interband ones. Interband couplings are therefore commonly present in multicomponent electronic systems, whatever the microscopic origin, orbital or geometrical. Our considered multicomponent system incorporates the above described physics and does not depend on a particular origin of the pairing interaction.\\

The manuscript is organized as follows. Section II details the physical system under study and the mean-field theoretical approach for evaluating the superconducting properties. In Section III we report the results and discussion of our analysis with regard to the superfluid phase stiffness and the BKT transition temperature. The conclusions of our work are given in Section IV.

\section{System and Model}
 Consider a single-band 2D tight-binding model on the square lattice with energy dispersion,
\begin{equation}
    \epsilon(\textbf{k}) = -2 t [\cos{(k_{x}a)}+\cos{(k_{y}a)}]
\end{equation}
where $\textit{a}$ is the lattice constant. The nearest neighbour hopping parameter $\textit{t}$ is set to be 0.1 $eV$ and the wave-vectors belong to the first Brillouin zone $-\frac{\pi}{a} \leq k_{x,y} \leq \frac{\pi}{a}$. In this article, we use the units $\hbar$ = $\textit{a}$ = $k_{B}$ = 1. In this work, for simplicity, we don't consider the competition of superconductivity with charge order close to half filling because a small next-nearest neighbour hopping will circumvent the competition and, in general, the charge order due to nesting configurations of the Fermi surface is not commonly observed \cite{Fermi_nesting}. \\

The energy dispersion of the considered 2D quasi-flat band has a bandwidth of 0.8 eV. The chemical potential is positioned at the center of the band, coinciding with the location of a pronounced van Hove singularity in the density of states (DOS). Here, we intend strong or weak van Hove singularities, referring to the values of the DOS close to the singularities in a comparative manner.\\
The effective pairing interaction  is approximated by a separable potential $V(\boldsymbol{k},\boldsymbol{k^{'}})$ with an energy cutoff $\omega_{0}$  and it is given by:
\begin{center}
$V(\boldsymbol{k},\boldsymbol{k^{'}})$ = $-V_{0}\Theta(\omega_{0}-|\xi_{\boldsymbol{k}}|)\Theta(\omega_{0}-|\xi_{\boldsymbol{k^{'}}}|).$  \\
\end{center}
$V_{0} > 0$ is the strength of the attractive potential. We solve numerically the self-consistent BCS mean-field gap equation at a finite temperature Eq.~\eqref{gap} coupled with the density equation Eq.~\eqref{n_density}.
 \begin{equation}\label{gap}
 \begin{split}
  \Delta(\textbf{k}) = & -\frac{1}{\Omega} \sum_{\boldsymbol{k^{'}} } \Biggr[  V(\boldsymbol{k},\boldsymbol{k^{'}})  \frac{ \tanh{ \frac{E({\boldsymbol{k^{'}}})}{2T}} \Delta( \boldsymbol{k^{'}} ) } {2\sqrt{\xi(\boldsymbol{k^{'}})^{2} + \Delta(\boldsymbol{k^{'}})^{2} } }    \Biggr],\
\end{split}
\end{equation}

\begin{equation}\label{n_density}
\begin{split}
 n = & \frac{2}{\Omega}\sum_{\boldsymbol{k}}  \Biggr[  \frac{1}{2}\Bigl( 1-  \frac{ \xi(\boldsymbol{k})}{  \sqrt{\xi(\boldsymbol{k})^{2} +  \Delta(\boldsymbol{k})^{2} } }  \Bigl)f(-E({\boldsymbol{k}}))\\    
    &+\frac{1}{2} \Bigl( 1+  \frac{ \xi(\boldsymbol{k})}{  \sqrt{\xi(\boldsymbol{k})^{2} + \Delta(\boldsymbol{k})^{2} } }   \Bigl)f(E({\boldsymbol{k}})) \Biggr].
 \end{split}
\end{equation}
The factor 2 in Eq.~\eqref{n_density} is due to spin degeneracy. $E_{\boldsymbol{k}} = \sqrt{\xi(\boldsymbol{k})^{2}+ \Delta(\boldsymbol{k})^{2}}$  is the dispersion of single-particle excitations in the superconducting state. $\Omega$ is the area occupied by the 2D system.
$\xi(\boldsymbol{k}) =  \epsilon(\boldsymbol{k})-\mu$ is the dispersion relation for the electronic band with respect to the chemical potential ($\mu$). In the single-band case, the dimensionless coupling constant for a pairing of two electrons is defined as ${\lambda}$ = $N(E_{0})V_{0}$  where $N(E_{0})$ is the density of states at the bottom of the band. $f(x) =[1 + \exp(x/T)]^{-1}$   is the Fermi-Dirac distribution function.\\
We point out that all the coupling configurations considered in this work led to solutions of the gap equations in Eq.~\eqref{gap} which are global minima of the free energy and thus stable physical solutions for the two-gap superconducting state, as discussed in Ref.~\onlinecite{subolo}.
The BKT transition temperature $T_{BKT}$ can
be evaluated by self-consistently solving the Kosterltiz-Thouless condition \cite{kt}:
\begin{equation}\label{4}
    T_{BKT} =  \frac{\pi}{2} \rho_{s}(\Delta(T_{BKT}),\mu(T_{BKT}),T_{BKT}).
\end{equation}
where $\rho_{s}(T)$ is the superfluid stiffness, $\Delta$ and $\mu$ are temperature dependent and given by the solutions of gap Eq.~\eqref{gap} and density Eq.~\eqref{n_density}. The superconducting gap has the same energy cutoff and the same wave-vector dependence of the interaction:

\begin{equation}
    \Delta(\boldsymbol{k}) = \Delta \Theta(\omega_{0}-|\xi_{\boldsymbol{k}}|).  
\end{equation}
Superfluid stiffness at  BCS mean-field level for a generic band and at a finite temperature $T$ is \cite{lara, Walter}
\begin{equation}\label{rho_s}
\begin{split}
     \rho_{s}(T) & = \frac{1}{4}\int \frac{d^{2}k}{(2\pi)^{2}} \frac{\partial^{2}\epsilon(\boldsymbol{k}) }{\partial k_{\alpha}^{2}} \left[ 1 - \frac{\xi({\boldsymbol{k}}) }{E_{\boldsymbol{k}}}\tanh\left({\frac{E_{\boldsymbol{k}}}{2T}}\right) \right] \\
 &+ \frac{1}{2} \int \frac{d^{2}k}{(2\pi)^{2}} \left( \frac{\partial\epsilon(\boldsymbol{k})}{\partial k_{\alpha}}  \right)^{2}\frac{\partial f (E_{\boldsymbol{k}})}{\partial E_{\boldsymbol{k}}}.
 \end{split}
\end{equation}
Where $k_{\alpha} = k_{x,y}$. Due to the tetragonal symmetry of the system $\rho_{x,x}$ = $\rho_{y,y}$ = $\rho_{s}.$\\

The two-band 2D systems in which one band is deep and the other is quasi-flat in nature is then considered.
    \begin{equation*}
    \begin{split}
    \epsilon_{i}(\boldsymbol{k}) & =  -2 t_{i} [\cos(k_{x}a)+\cos(k_{y}a)] + E_{g,i}.
    \end{split}
\end{equation*}

\begin{figure}[t]
\centering
\includegraphics[width=8.5cm,height = 5.5cm]{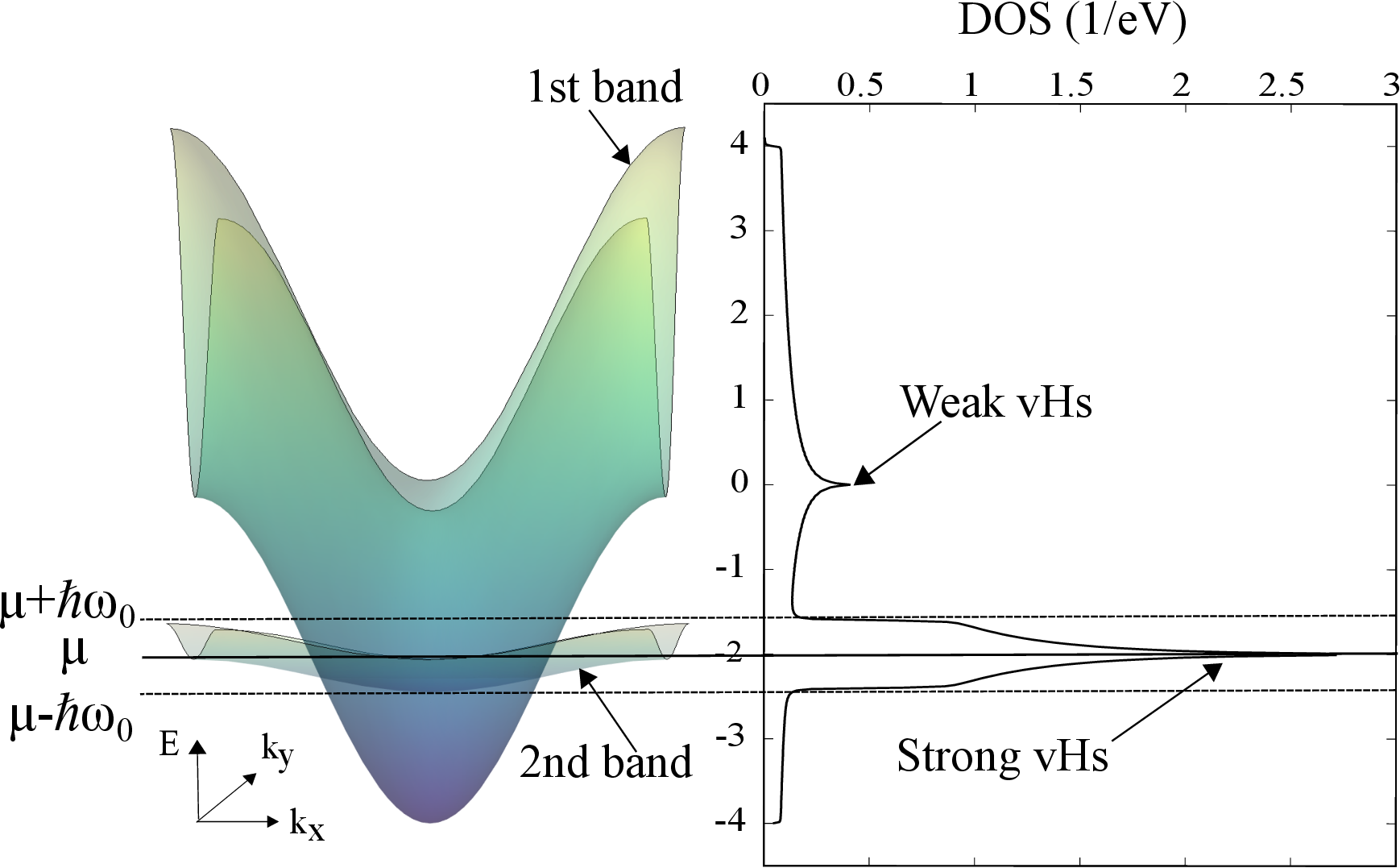}
 \caption{Band 1 corresponds to the deep band with a weak van Hove singularity at E = 0 and band 2 corresponds to the quasi-flat band having a strong van Hove singularity at E = $E_{g,2}$ = -2$t_{1}$.}
    \label{2}
\end{figure}

The index $\textit{i}$ = 1,2 labels the bands: $\textit{i}$ = 1 denotes the upper deep band having the nearest neighbour hopping parameter $t_1$ = 1.0 $eV$. $\textit{i}$ = 2 denotes the lower band which has the characteristic of quasi-flat band with $t_{2}$= 0.1$t_{1}$. The quasi-flat band is shifted below the center of the deep band by the energy $E_{g,2}$ = -2$t_{1}$ and no energy shift for the deep band is considered $E_{g,1}$ = 0.  $E_{g}$ = $1.6 t_{1}$ is the energy difference between the bottom of the two bands.\\
The effective multichannel pairing interaction is approximated using a separable potential, which is given by:
\begin{center}
$V_{ij}(\boldsymbol{k},\boldsymbol{k^{'}})$ = $-V^{0}_{ij}\Theta(\omega_{0}-|\xi_{i}(\boldsymbol{k})|)\Theta(\omega_{0}-|\xi_{j}(\boldsymbol{k^{'}})| ). $  \\
\end{center}
$V_{i,j}^{0}$ is the strength of the potential and $\textit{i, j}$ label the bands. $V^{0}_{ij}$ = $V^{0}_{ji}$. $V^{0}_{11}$ and $V^{0}_{22}$ are the strengths of the intraband pairing interactions. $V^{0}_{12}$ = $V^{0}_{21}$ are the strength of the pair-exchange interactions. We consider the same energy cutoff $\omega_{0}$ of the interaction for the pair-exchange and intraband pairing terms. 
The considered energy cutoff spans the entire spectrum of the quasi-flat band, aiming to encompass all available wave-vector states for pairing. Because of the flatness in energy of the quasi-flat band, with bandwidths of the order of 10-100 meV, the typical energy scale for the boson mediators of the pairing, of phononic or electronic origin, is of the order of the quasi-flat band bandwidth, but it results smaller than the bandwidth of the deep dispersive band. Consequently, there exists coexistence between local (BEC-like) pairs forming from the quasi-flat band and more extended Cooper (BCS-like) pairs emerging from the deep band. The experimental observation of BCS-BEC crossover in iron-based or other 2D superconductors \cite{ Iwasa, shibauchi, shibauchi2,crossover} demonstrates that all the electronic states of one band collapse in forming the BEC condensate of local pairs, a strong coupling regime that requires the pairing energy cutoff to be at least of the order of the bandwidth. The regime corresponding to small energy cutoffs with respect to the bandwidth is instead typical of BCS regimes of conventional superconductivity.
Another realization of pairing energy cutoffs, due to phonon exchanges, of the order of the electronic bandwidths, can be found in intercalated fullerene superconductors, in which the strong electronic correlations act in shrinking the bands (enhancing the effective mass) in the presence of sizable energy of the relevant phononic model \cite{m.capone}.  It is therefore a quite common configuration to be found in multiband and/or strongly correlated superconductors. For concentric or non-concentric Fermi surfaces, once the effective pairings are
parametrized in terms of intraband and interband channels, the structure of the multiband self-consistent gap equations does not depend on the specific relative position in the BZ of the Fermi surfaces.
In the effective model, BCS or GL, even starting from displaced Fermi surfaces in the BZ, once effective coupling parameters have been introduced for the Cooper pairing of multiple types of electrons, the displacement disappears from the self consistent gap or $T_{c}$ equations, with
or without fluctuations \cite{two_patch_1}.  In this study, we utilize a mean-field approach that is based on the two-band extension of the mean-field BCS theory \cite{suhl,2_perali} at finite temperature. The resulting BCS equations for the two superconducting gaps are coupled with an equation for the total density of the system.
The equations for the superconducting gaps $\Delta_{1}(\textbf{k})$ and $\Delta_{2}(\textbf{k})$ read:
 \begin{equation}\label{gap_1}
 \begin{split}
  \Delta_{1}(\textbf{k}) = & -\frac{1}{\Omega} \sum_{\boldsymbol{k^{'}} } \Biggr[  V_{11}(\boldsymbol{k},\boldsymbol{k^{'}})  \frac{ \tanh{\frac{E_{1}({\boldsymbol{k^{'}}})}{2T}} \Delta_{1}( \boldsymbol{k^{'}} )} {2\sqrt{\xi_{1}(\boldsymbol{k^{'}})^{2} + \Delta_{1}(\boldsymbol{k^{'}})^{2}}}  \\
    & +   V_{12}(\boldsymbol{k},\boldsymbol{k^{'}})  \frac{\tanh{\frac{E_{2}({\boldsymbol{k^{'}}})}{2T}} \Delta_{2}( \boldsymbol{k^{'}} )   } {2\sqrt{\xi_{2}(\boldsymbol{k^{'}})^{2} + \Delta_{2}(\boldsymbol{k^{'}})^{2} } }  \Biggr],\\
\end{split}
\end{equation}
 \begin{equation}\label{gap_2}
 \begin{split}
  \Delta_{2}(\textbf{k}) = & -\frac{1}{\Omega} \sum_{k^{'} } \Biggr[  V_{21}(\boldsymbol{k},\boldsymbol{k^{'}})  \frac{ \tanh{\frac{E_{1}(\boldsymbol{k^{'}})}{2T}} \Delta_{1}( \boldsymbol{k^{'}} ) } {2\sqrt{\xi_{1}(\boldsymbol{k^{'}})^{2} + \Delta_{1}(\boldsymbol{k^{'}})^{2} } }  \\
    & +   V_{22}(\boldsymbol{k},\boldsymbol{k^{'}})  \frac{ \tanh{ \frac{E_{2}(\boldsymbol{k^{'}})} {2T}}  \Delta_{2}( \boldsymbol{k^{'}} )} {2\sqrt{\xi_{2}(\boldsymbol{k^{'}})^{2} + \Delta_{2}(\boldsymbol{k^{'}})^{2} } }  \Biggr],\\
\end{split}
 \end{equation}
where $\mu$ is the chemical potential fixed to be the same for both bands. $\xi_{i}(\boldsymbol{k}) =  \epsilon_{i}(\boldsymbol{k})- \mu$ is the energy dispersion of the bands with respect to the chemical potential. The dimensionless couplings $\lambda_{i,j}$  are defined as $\lambda_{i,j}$ = $V_{i,j}^{0}N(E_{0})$, $N(E_{0})$ is the density of states at the bottom of the deep band. The total density of the two-band system is fixed, and it is given by the sum of the individual densities of the bands, $n_{tot} = n_{1} + n_{2}.$ The fermionic density $n_{i}$ in the $\textit{i}^{th}$ band is given by:
\begin{equation}\label{n_1}
\begin{split}
 n_{i} = & \frac{2}{\Omega}\sum_{\boldsymbol{k}}  \Biggr[ \frac{1}{2} \Bigl( 1-  \frac{ \xi_{i}(\boldsymbol{k})}{  \sqrt{\xi_{i}(\boldsymbol{k})^{2} +  \Delta_{i}(\boldsymbol{k})^{2} } }  \Bigl)f(-E_{i}(\boldsymbol{k}))\\    
    &+ \frac{1}{2} \Bigl( 1+  \frac{ \xi_{i}(\boldsymbol{k})}{  \sqrt{\xi_{i}(\boldsymbol{k})^{2} + \Delta_{i}(\boldsymbol{k})^{2} } }   \Bigl)f(E_{i}(\boldsymbol{k})) \Biggr].
 \end{split}
\end{equation}
 The system of equations \eqref{gap_1}, \eqref{gap_2} and \eqref{n_1} with the constraint $n_{tot}$ = $n_{1}$ + $n_{2}$ is solved self consistently. The resulting values of the superconducting gaps $\Delta_{1}$, $\Delta_{2}$ and chemical potential $\mu$ are used to calculate the total superfluid stiffness of the system \cite{lara,chubukov}:
\begin{equation}\label{total_rho}
    \rho^{tot}_{s} = \rho^{deep}_{1} + \rho^{qf}_{2},
\end{equation}
where $\rho^{deep}_{1} $ and $\rho^{qf}_{2} $ are the superfluid stiffness of the deep band and quasi-flat band condensates, respectively. In the absence of the cross-pairing interaction \cite{paredes} in our two-band system, it becomes possible to express the total superfluid stiffness as the sum of the individual stiffness contributions from each band-condensate \cite{lara, chubukov}. In this work, we neglect only the cross-pair interactions, i.e. the possibility to form Cooper pairs with one electron in one band and the second electron in the other band. Generally, incorporating cross-band pairing results in a complex hybridization of excitation spectra in the superconducting state. This introduces a competition between the formation of Cooper pairs within the same band and across different bands, a problem discussed in detail in Ref. \cite{paredes}. In presence of significant cross-pairing, specific spectroscopic features (which are not commonly observed in multiband superconductors) are expected \cite{paredes}. In the present work, we limit our analysis to a regime of parameters with negligible cross-pairing amplitudes, favoring conventional multiband superconductivity with only pair-exchange effects. Moreover, the superfluid phase stiffness in a two-band system having sizable cross-pairing is no longer a straightforward sum of individual band stiffnesses.
 The BKT transition temperature $T_{BKT}$ for the coupled two-band 2D superconducting system can be evaluated by self-consistently solving the Kosterltiz-Thouless condition \cite{kt}:
\begin{equation}\label{4.1}
    T_{BKT} =  \frac{\pi}{2} \rho^{tot}_{s}(\Delta_{1}(T_{BKT}),\Delta_{2}(T_{BKT}),\mu(T_{BKT}),T_{BKT}).
\end{equation}
In the next section, we report the results and discussion of single-band and two-band cases.

\section{RESULTS AND DISCUSSION}
\subsection{Single band case}
In the single quasi-flat band tight-binding model in 2D, at zero temperature, the chemical potential is fixed around the strong van Hove singularity of the quasi-flat band. Eq.~\eqref{rho_s} becomes
\begin{equation*}
\begin{split}
    \rho^{qf}_{s}(T=0) & = \frac{1}{4}\int \frac{d^{2}k}{(2\pi)^{2}} \frac{\partial^{2}\epsilon(\boldsymbol{k})}{\partial k_{\alpha}^{2}}
\left[ 1 - \frac{\xi(\boldsymbol{k})}{\sqrt{\xi(\boldsymbol{k})^{2} + \Delta_{qf}(\boldsymbol{k})^{2}} } \right]. \\
\end{split}
\end{equation*}
 Performing a large  gap expansion $\Delta_{qf}$/t $>>$ 1,
\begin{equation}\label{semi_single}
\rho^{qf}_{s}(T=0)  \approx     \rho^{qf}_{0} - \frac{\rho^{qf}_{1}}{\Delta_{qf}} + \frac{\rho^{qf}_{2}}{\Delta^{3}_{qf}} +...
\end{equation}
where,
\begin{equation*}
\rho^{qf}_{0} =  \frac{1}{4}\int \frac{d^{2}k}{(2\pi)^{2}} \frac{\partial^{2}\epsilon(\boldsymbol{k})}{\partial k_{\alpha}^{2}},
\end{equation*}

\begin{equation*}
\rho^{qf}_{1}  =  \frac{1}{4}\int \frac{d^{2}k}{(2\pi)^{2}} \frac{\partial^{2}\epsilon(\boldsymbol{k})}{\partial k_{\alpha}^{2}}\xi(\boldsymbol{k}) , 
\end{equation*}

\begin{equation*}
\rho^{qf}_{2}  =  \frac{1}{4}\int \frac{d^{2}k}{(2\pi)^{2}} \frac{\partial^{2}\epsilon(\boldsymbol{k})}{\partial k_{\alpha}^{2}} \frac{ \xi(\boldsymbol{k})^{3}}{2}.   
\end{equation*}

\begin{figure}[t]
\includegraphics[width= \linewidth]{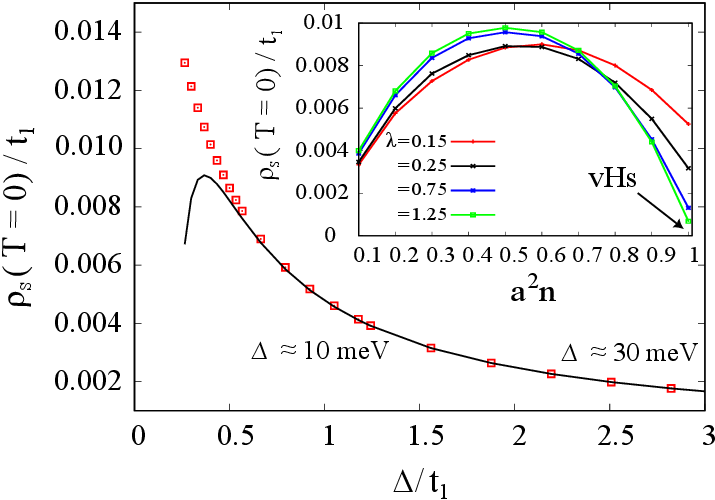}
\caption{The superfluid stiffness (open squares) at zero temperature as a function of the superconducting gap $\Delta$/$t_{1}$ for a given cutoff $\omega_{0} = 4t$. In comparison with semi-analytical stiffness (solid line) of Eq.~\eqref{semi_single}.  Inset: The superfluid stiffness at zero temperature is reported as a function of density for different couplings $\lambda$ = 0.15 (red), 0.25 (black), 0.75 (blue), 1.25 (green). }
    \label{fig_2}
\end{figure}

In general, quasi-flat bands are having the characteristic of higher effective mass $m^{*}$ and superconducting gap $\Delta_{qf}$, which is inversely related to the superfluid stiffness by Eq.~\eqref{semi_single}. As a result, the Berezinskii-Kosterlitz-Thouless transition temperature subsequently dropped. The magnitude of this effect is expected to increase when a strong van Hove singularity (vHs) is approached. The presence of a strong vHs in the density of states enhances the mean-field critical temperature $T_{c}$ \cite{Hirsch,Mark}, However, on the other hand, it suppresses the superfluid stiffness $\rho_{s}$ and consequently the BKT transition temperature $T_{BKT}$, due to a larger superconducting gap. In Fig.~\ref{fig_2}, the superfluid stiffness at zero temperature is reported as a function of the superconducting gap $\Delta$ in units of $t_{1}$. The deviation of the numerical results from the semi-analytical results around a superconducting gap value of 0.75 is only a 5$\%$. However, when superconducting gap exceeds the value of 1, the numerical results for superfluid stiffness match very well with the semi-analytical results as provided by the Eq.~\eqref{semi_single}.

\begin{figure}[t]
\includegraphics[width=\linewidth]{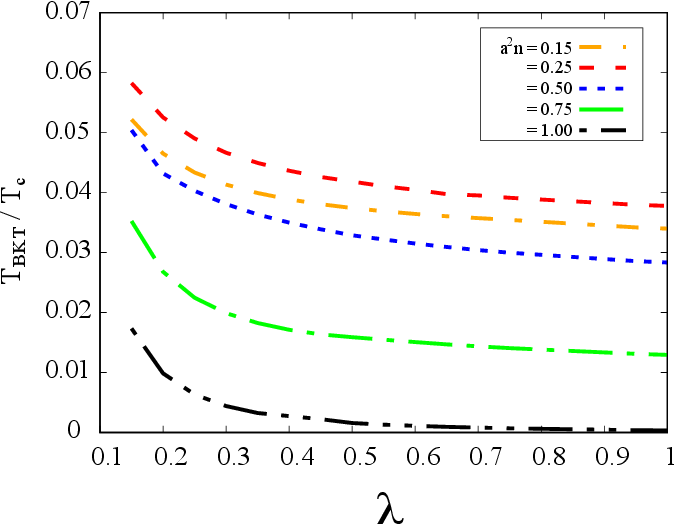}
 \caption{$T_{BKT}/T_{c}$ against the coupling strength $\lambda$ of the 2D quasi-flat band for different densities and given energy cutoff $\omega_{0} = 4t$. }
 \label{fig_3}
\end{figure}

At zero temperature, as the system approaches the strong coupling regime, the superconducting gap increases, resulting in a further reduction of the superfluid stiffness. Regardless of the band geometry, the superconducting gap becomes the dominant factor affecting the superfluid stiffness in the strong coupling regime. This is due to the fact that the superconducting gap is directly related to the pairing strength of electrons. Therefore, in the strong coupling regime, the superconducting gap is the primary determinant of the superfluid stiffness of the system. In the inset of Fig.~\ref{fig_2}, the superfluid stiffness at zero temperature shows the non-monotonic behaviors with respect to the number density and achieves its peak magnitude when the pairing strength is higher, particularly around the quarter filling. However, it undergoes a more substantial suppression in the vicinity of the van Hove singularity. At zero temperature, the superfluid stiffness at the bottom of the tight-binding band can be approximated by the parabolic band model, expressed as $\rho_{s}$ $\approx$ $\frac{n}{m^{*}}$. In this regime, the superfluid stiffness is influenced by the effective mass.

 \begin{figure}[b]
\includegraphics[width=\linewidth]{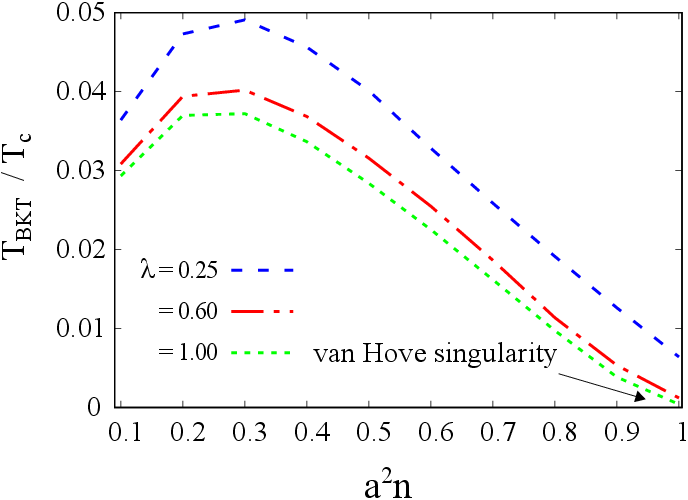}
 \caption{$T_{BKT}/T_{c}$, as a function of density for $\lambda$ = 0.25 (blue), 0.60 (red), 1.00 (green). }
 \label{fig_3.01}
\end{figure}

 In Fig.~\ref{fig_3}, we report $T_{BKT}/T_{c}$ versus the coupling strength for different values of electronic densities. In the weak coupling region,  $T_{BKT}/T_{c}$ exhibits  relatively a high value. In the BCS weak coupling limit, the superconducting gap $\Delta$ is exponentially suppressed for small values of the coupling strength $\lambda$. As the coupling strength increases and approaches the strong coupling regime, the gap becomes larger, which in turn suppresses more the superfluid stiffness and $T_{BKT}$ as described by Eq.~\eqref{semi_single}. In the case of half filling  $a^{2}n$ = 1.0, $T_{BKT}$/$T_{c}$ is notably suppressed because of the presence of the van Hove singularity \cite{T.Paiva}. As we increase the density of electrons from 0.15 to 0.25, $T_{BKT}$ initially increases. However, as we further increase the density, the superconducting gap dominates more, which leading to a suppression of the $T_{BKT}$. In order to screen the suppression and, hence, amplify $T_{BKT}$, particularly around the half-filling region, multiband superconductivity can be efficiently employed, as investigated below. In Fig. \ref{fig_3.01}, $T_{BKT}$/$T_{c}$ is plotted against the electron density for given coupling strength. At a state of half-filling, the ratio $T_{BKT}$/$T_{c}$ undergoes significant suppression, primarily due to the presence of the van Hove singularity. Irrespective of the coupling strength, it is noteworthy that the $T_{BKT}$/$T_{c}$ ratio attains its maximum value around quarter-filling.

 \begin{figure}[b]
\includegraphics[width= \linewidth]{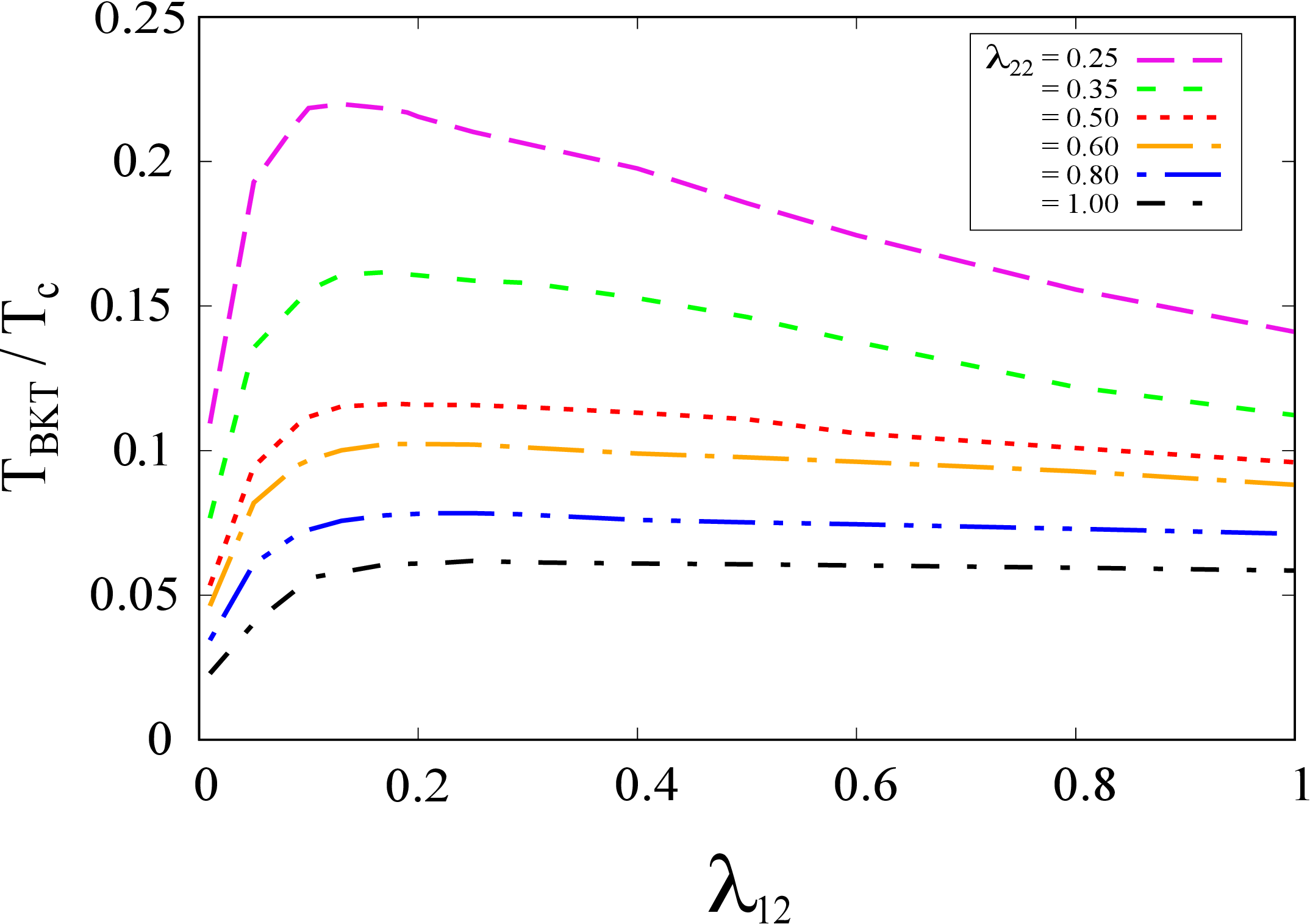}
 \caption{$T_{BKT}/T_{c}$ of the two-band system, as a function of the pair-exchange coupling strength ${\lambda_{12}}$ for different values of intraband coupling ${\lambda_{22}}$ at given $\omega_{0}$ = 4.0$t_{2}.$ }
    \label{fig_5}
\end{figure}
\subsection{Two-band case}
To address the issue of suppressed superfluid stiffness in the quasi-flat band 2D tight-binding system, we coupled the quasi-flat band with the 2D tight-binding deep dispersive band, which has a weak coupling strength and a broad van Hove singularity in the DOS.
The coupling strength for the deep band is fixed to be 0.25 throughout the calculation. In many different multiband superconductors, the resulting superconducting gaps are well separated in energy, with distinct features measured in spectroscopic experiments that can be attributed to different gaps. When considering pairing mediated by phonons or by other bosonic fields, the coupling strength between the bosonic mediator and the electrons forming the Cooper pairs depends strongly on the specific structure of the wave functions and, hence, of the electronic bands. Therefore, having different paring strengths in different bands is a standard situation in multiband systems, while equal pairings can be associated with quasi-degeneracy of the electronic structure. A prototype example would be MgB$_{2}$ or several iron-based superconductors. Multiband superconductors with a coherent mixture of condensates in the BCS regime (deep band) and in the BEC regime (quasi-flat band) offer a promising route to higher critical temperatures. In systems with quasi-flat bands, the flatness of the bandwidth can enhance the coupling strength through the large density of states in the flat regions of the band structure in the BZ, leading to the emergence of strongly correlated states and strongly-coupled superconductivity, characterized by the formation of short-sized pairs typical of the BEC regime. Thus, prioritizing stronger coupling in the quasi-flat band over the deep band is more favorable. 
\begin{figure}[t]
\includegraphics[width=\linewidth]{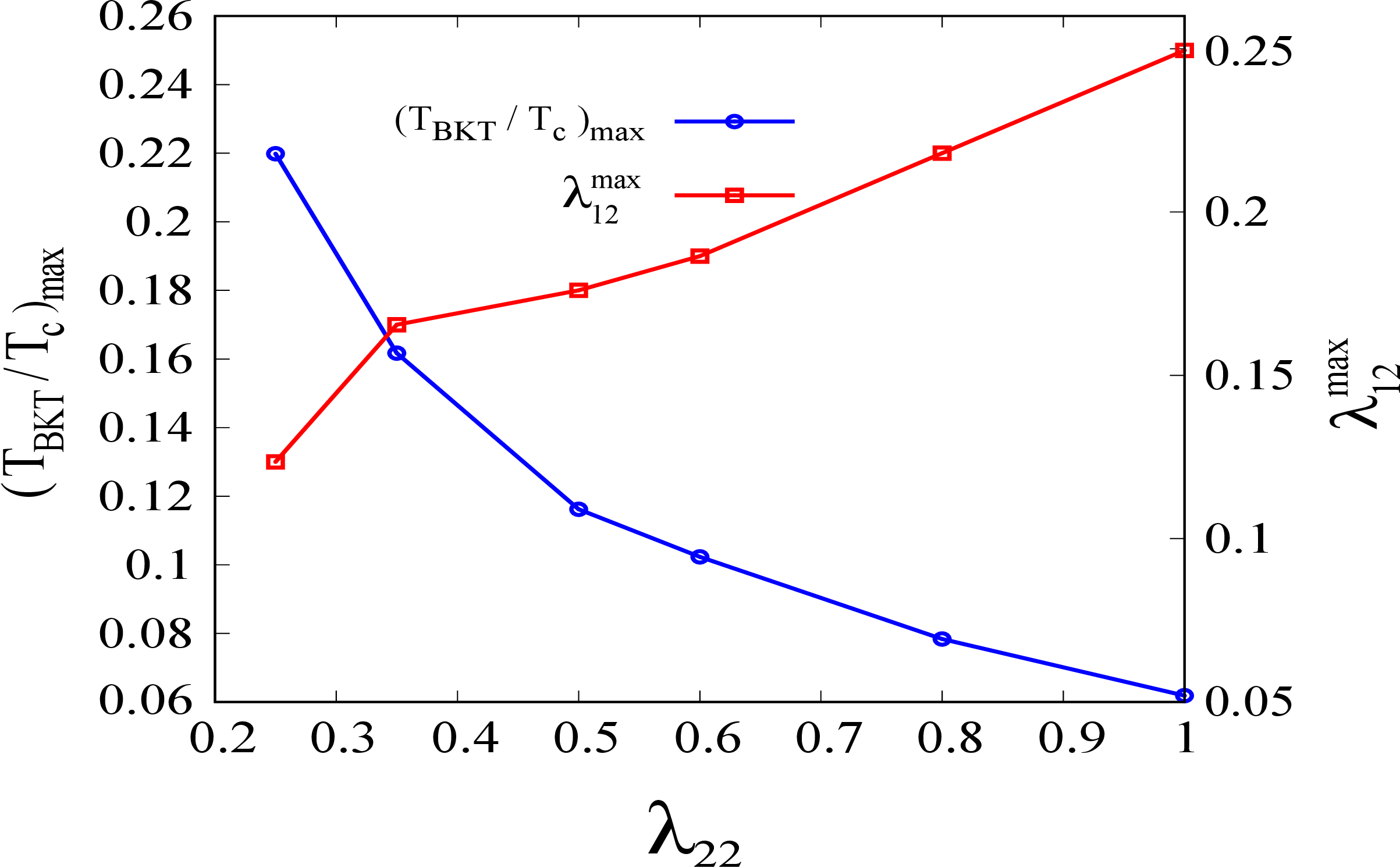}
 \caption{The maximum value of  $(T_{BKT}/T_{c})_{max}$ and its corresponding pair-exchange coupling constant $\lambda^{max}_{12}$ between the band condensates, as a function of the intraband coupling constant $\lambda_{22}$ of the quasi-flat band. }
 \label{fig_3.1}
\end{figure}
Same physics can arise in the presence of anti-adiabatic effects beyond Migdal, that can amplify electron-phonon coupling in the case of quasi-flat bands. At zero temperature, it is not possible to perform a large gap expansion of superfluid stiffness for a deep band, as the coupling constant ($\lambda_{11}$) of the deep band is fixed in the weak coupling regime. However, for $\Delta_{deep}$/$t_{1}$ $<$ 1, $\rho^{deep}_{1} $ can be approximated as  \\
\begin{equation}\label{equ_12}
\begin{split}
\rho^{deep}_{1}(T=0) \approx &    \rho^{deep}\left(\Delta_{deep}=0\right) +\\
  & \frac{1}{4} N_{0}(\mu)  \biggr[\Delta_{deep}- \frac{\Delta^{2}_{deep}}{2\omega_{0}} \biggr] \sum_{k}\frac{\partial^{2}\epsilon(\boldsymbol{k})}{\partial k_{\alpha}^{2}},
\end{split}
\end{equation}
where $N_{0}(\mu)$ is the density of the states at the chemical potential. Since the superfluid stiffness of the deep band ($\rho^{deep}_{1}$) is directly related to the superconducting gap of a deep band which acts as a reservoir to enhance the total superfluid stiffness ({$\rho^{tot}_{s}$}) of the system given by Eq.~\eqref{total_rho}. The position in energy of the strong van Hove singularity originating from the quasi-flat band is not a relevant issue in our model and approach. Even when shifting the center of the quasi-flat band to the higher energies, the stabilization effect proposed in our work will not be significantly affected, with the only condition that the deep band should not be completely filled or completely empty of electrons, otherwise its contribution to the superconducting phase stiffness will vanish and such deep band will not be active for the superconductivity. \

 \begin{figure}[b]
\includegraphics[width=\linewidth ]{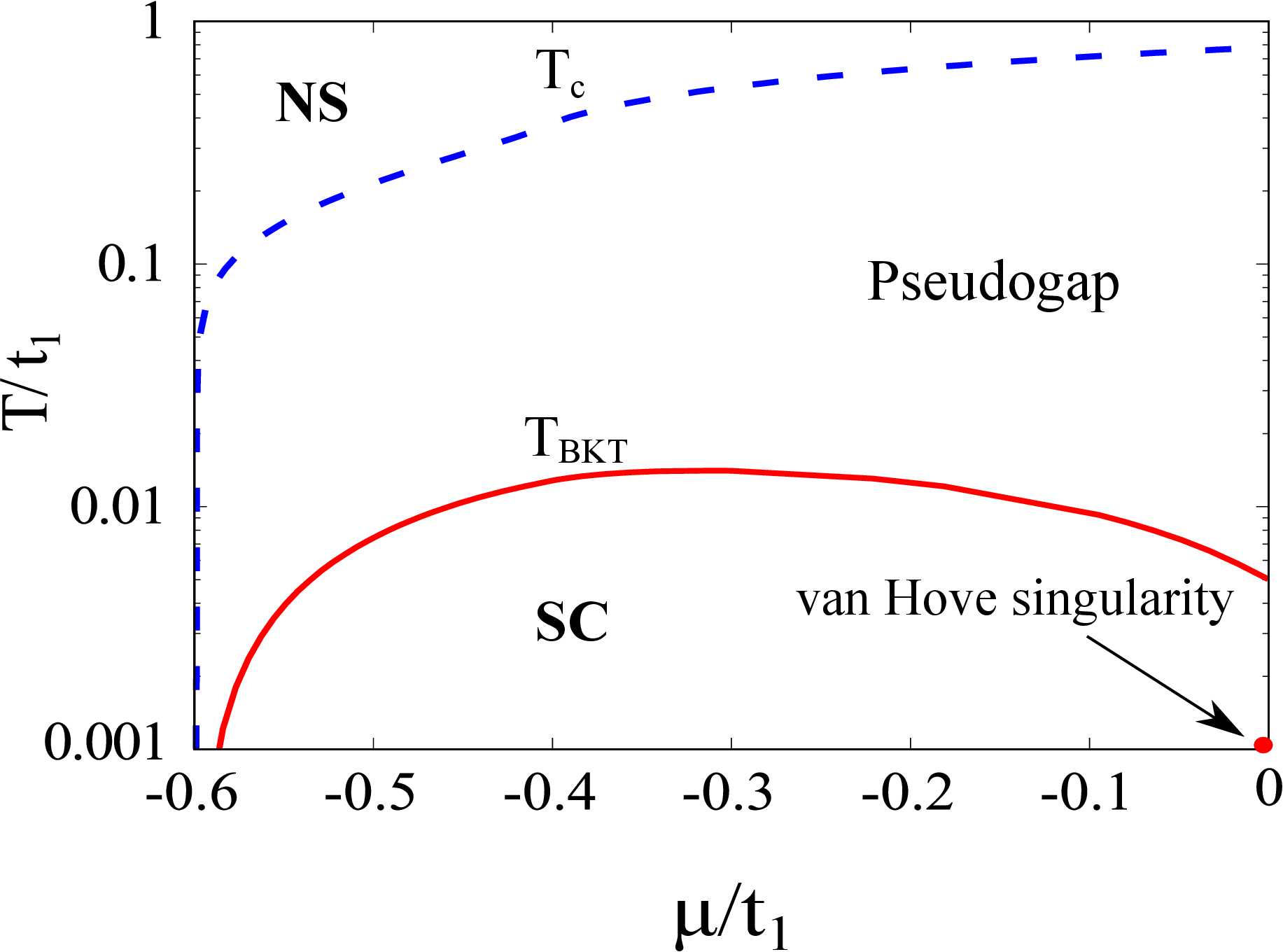}
 \caption{Mean-field critical temperature $T_{c}$ and BKT transition temperature $T_{BKT}$ as a function of the chemical potential, in the single-band case, for a given cutoff $\omega_{0}$ = 4$t_{2}$, and $\lambda$ = 0.25.}
    \label{fig_6}
\end{figure}

\begin{figure}[t]
\includegraphics[width= \linewidth ]{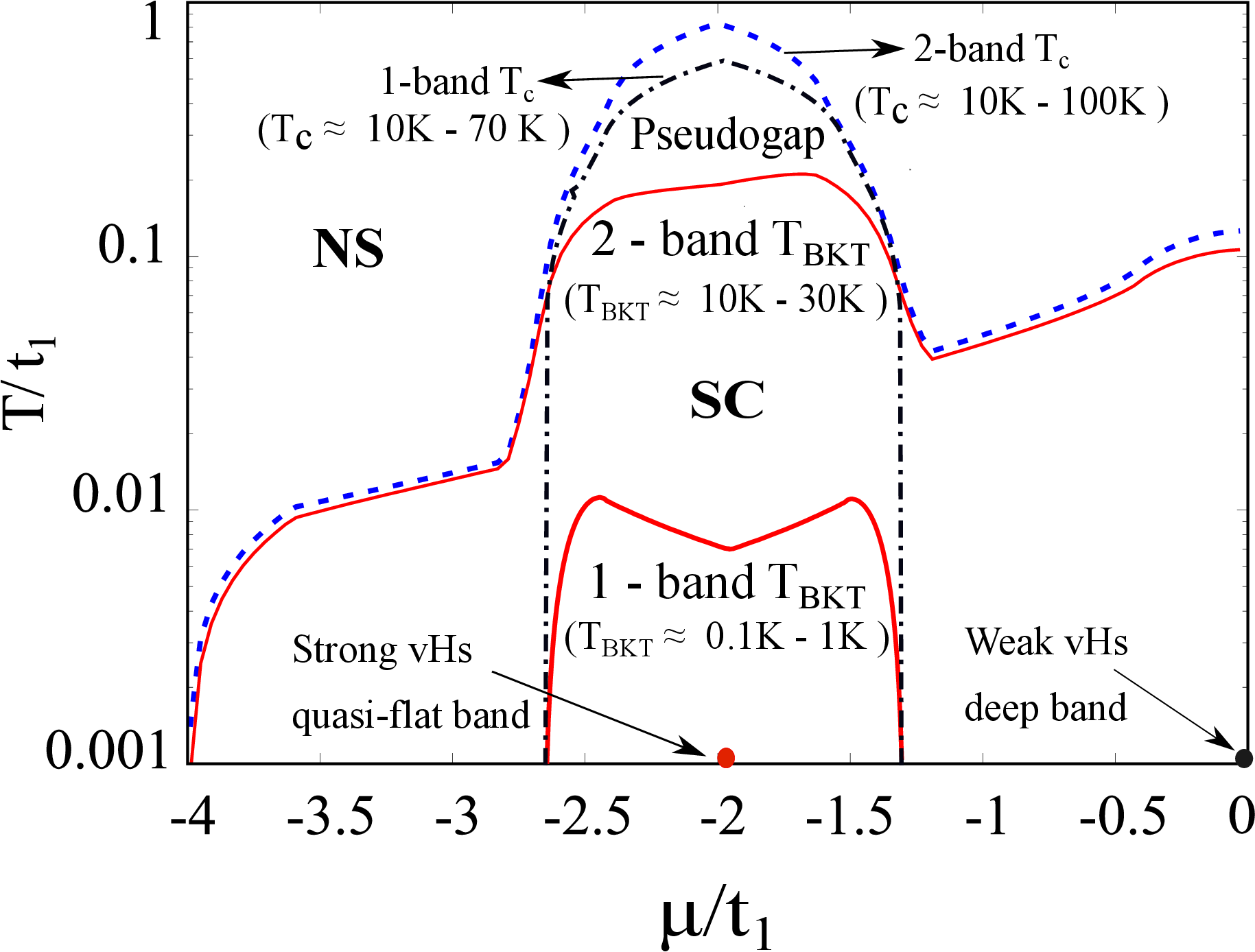}
 \caption{Mean-field critical temperature $T_c$ and Berezinskii–Kosterlitz–Thouless transition temperatures $T_{BKT}$ as a function of the chemical potential, in the two-band case, for a given cutoff energy $\omega_{0}$ = 4$t_{2}$, with coupling strengths $\lambda_{11}$ = 0.25, $\lambda_{22}$ = 0.25, $\lambda_{12}$ = $\lambda_{21}$ = 0.2. A single-band $T_{BKT}$ is plotted as well, to highlight the amplification between the single-band scenario and the two-band scenario. }
    \label{fig_7}
\end{figure}

In the flat band and quasi-flat band systems, the superfluid stiffness and BKT transition temperature are negligible due to band geometry. But recent research has indicated that incorporating quantum geometric effects in multiorbital band models could result in a finite BKT transition temperature \cite{torma_multi,torma_nature,band_geometry, Torma_2017,QG_review,revisit, fese_geometric}. The presence of quantum geometry is crucial for achieving stable supercurrent and superfluidity in quasi(flat) bands, particularly if the bands possess non-trivial topological quantum geometric properties. These properties are determined by the overlaps between the eigenstates in a band and are described mathematically by the quantum geometric tensor, with the real component known as the quantum metric. While the quantum metric has been associated with the superfluid stiffness in the isolated flat band limit \cite{torma_nature,torma_multi,Torma_2017,s.peotta,band_geometry,revisit, QG_review,  
 fese_geometric,Rossi}, our article considers a two-band model in which one quasi-flat band is located below the center of the deep band, indicating that we are not in the isolated flat band limit. Furthermore, we do not consider any non-trivial topological bands, which ultimately leads us to conclude that the superfluid stiffness is not influenced by quantum geometric contributions \cite{torma_multi}. Nonetheless, due to the band geometry, the single-band superconducting system in the quasi-flat band regime exhibits a very small but non-zero conventional superfluid stiffness.\\

In our two-band model, the chemical potential is fixed around the vicinity of the van Hove singularity of the quasi-flat band. This region is characterized by a high suppression of the superfluid stiffness and the opening of a pseudogap, which requires the utmost care in the analysis. The intraband and pair-exchanging coupling strengths and the chemical potential primarily determine the mean-field critical temperature in the two-band case and hence play a crucial role in determining the Berezinskii-Kosterlitz-Thouless transition temperature ($T_{BKT}$). To enhance the BKT transition temperature, the quasi-flat band is coupled with the deep band, where the latter practically acts as a passive band. We consider the intraband coupling strengths of the deep and quasi-flat bands, denoted by $\lambda_{11}$ and $\lambda_{22}$ respectively. As the deep band acts as a passive band, its intraband coupling is smaller than that of the quasi-flat band, i.e., $\lambda_{11} < \lambda_{22}$, while the pair-exchange coupling strengths are symmetric, i.e., $\lambda_{12} = \lambda_{21}$. Throughout the numerical calculation, we fix the deep band in the weak coupling regime $\lambda_{11}$= 0.25 and vary the  $\lambda_{22}$ ranging from 0.25 (weak coupling) to 1.0 (strong coupling). To determine the optimal conditions for achieving a higher BKT transition temperature ($T_{BKT}$), we vary the pair-exchange coupling strength, $\lambda_{12}$, from 0.1 to 1.0. To fix the chemical potential around the vicinity of the van Hove singularity of the quasi-flat band in the coupled system, we tune the total number density, $n_{tot}$. In Fig.~\ref{fig_5}, we plotted the ratio of $T_{BKT}/T_{c}$ for the different values of pair-exchange coupling strength ranging from 0.01 to 1.0. In the two-band model, in the weak coupling regime for the quasi-flat band with $\lambda_{22}=0.25$, the $T_{BKT}$ transition temperature is found to be approximately 20\% to 22\% of $T_{c}$ for the pair-exchange coupling regime, ranging from 0.1 to 0.4. Similarly, in contrast, the single quasi-flat band case, as depicted in Fig.~\ref{fig_3}, shows that for the coupling strength from 0.15 to 0.4 for  $a^{2}n=1.0$, the $T_{BKT}$ is only around 0.5\% to 1.5\% of $T_{c}$. This result suggests that coupling a deep band with the quasi-flat band significantly enhances the $T_{BKT}$ by approximately 10–100 times compared to the single-band case. From Eq.~\eqref{equ_12} the superfluid stiffness of the deep band is directly related to the superconducting gap $\Delta_{deep}$ which behaves like a reservoir of superfluid stiffness, screening the suppression of the superfluid stiffness. In Fig.~\ref{fig_3.1}, the maximum value of the  $T_{BKT}/T_{c}$ and its corresponding pair-exchange coupling $\lambda_{12}$ between the band condensates are plotted as a function of intraband coupling of the quasi-flat band. It shows that with the suitable choice of pair-exchange and intraband couplings, we can achieve the maximum enhancement of the $T_{BKT}$. The two-band system behaves like a single-band system as $\lambda_{12}$ is much larger than $\lambda_{11}$, $\lambda_{22}$, as a result there is a decline of the $T_{BKT}$ for very large values of $\lambda_{12}$. In Fig.~\ref{fig_6}, when the chemical potential reaches the van Hove singularity of the quasi-flat band, there is a huge opening of the pseudogap. In Fig.~\ref{fig_7}, for the two-band case, as the chemical potential approaches the strong van Hove singularity, there is a significant decrease in the pseudogap. This effect can be attributed to the screening-like mechanism that is associated with multiband superconductivity. 

\begin{figure}[b]
\centering
\includegraphics[width=9.0cm,height =7.0cm]{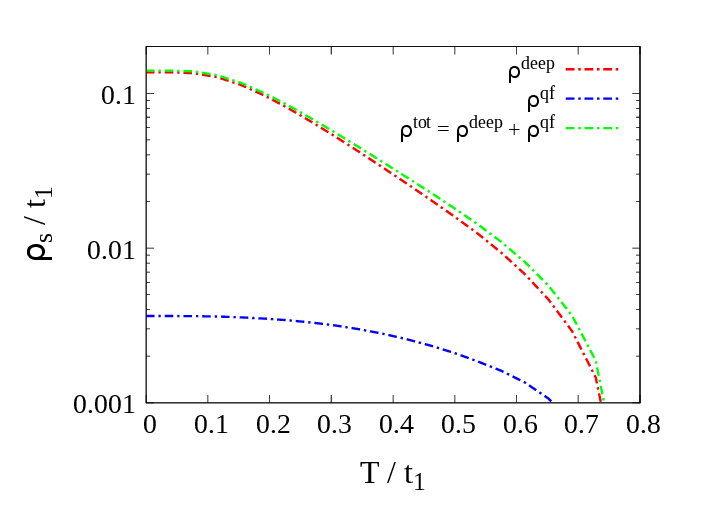}
 \caption{Superfluid stiffness as a function of the temperature for two band case with intraband coupling strength of deep and quasi flat band are 0.25 (red) and 0.25 (blue) respectively. The total superfluid stiffness is also shown (green), pair-exchange coupling strength $\lambda_{12}$ = 0.1. }
    \label{fig_8}
\end{figure}

\begin{figure}[h]
\centering
\includegraphics[width=\linewidth]{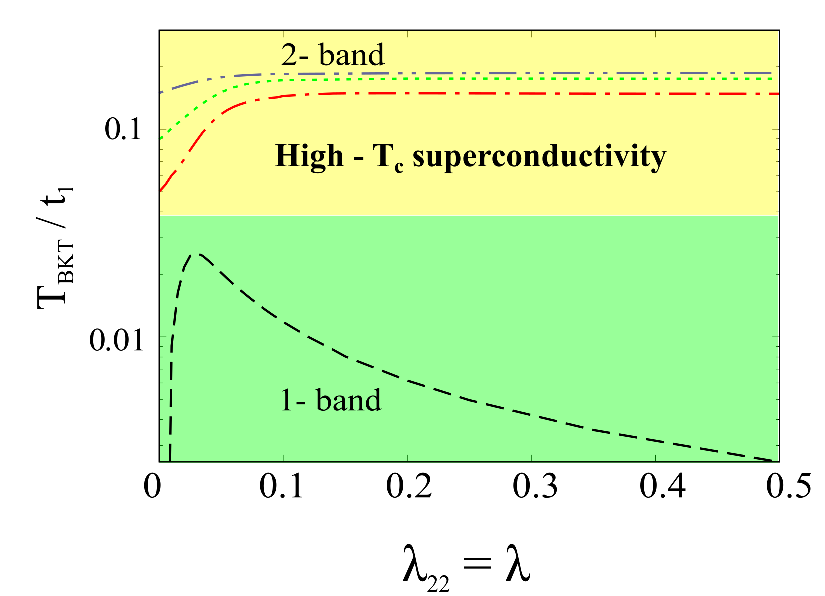}
 \caption{$T_{BKT}$/$t_{1}$ as a function of the pairing strength for both single-band and two-band cases. For the two-band case, pair-exchange coupling strengths considered are $\lambda_{12}$ = 0.05(red), 0.1(green), 0.2(blue). The intraband coupling strength for the deep band is $\lambda_{11}$ = 0.25. }
    \label{fig_9}
\end{figure}

For a more realistic description of monolayer FeSe, we opted for the hopping parameter $t_{1}$ = 10~meV and the coupling constants $\lambda_{11} = 0.25$, $\lambda_{22} = 0.25$, and $\lambda_{12} = 0.2$. Depending on the chemical potential, the resulting mean-field critical temperature $T_{c}$ for the single-band case is in the range 10-70~K, while for the two-band scenario, it is approximately 10-100~K.  The BKT transition temperature ($T_{BKT}$) for the single-band case falls within the range 0.1-1~K, while for the two-band case it spans 10~K to 30~K, closely matching the BKT temperature of around 24~K observed in FeSe monolayer superconductors as reported in Ref. \cite{zhang}. 
The experimental observation of the mean-field temperature for onset of pairing and opening of the pseudogap for the FeSe monolayer occurs around 55-65~K,  which coincides with the temperature range of 10 to 100~K extracted from parameter fits with our considered model. Moreover, in this parameter set, the superconducting gap for the two-band case ranges from 7 to 16 meV, reflecting values that are both reasonable and consistent  with the STM and ARPES studies reporting the superconducting gap value from 8 meV to 20 meV \cite{wang, He, zhang}.
In Fig. \ref{fig_8}, the superfluid phase stiffness is plotted as a function of temperature, showing how the presence of multigap/multi-band contributes to the enhancement of superfluid stiffness.
In Fig.~\ref{fig_9}, $T_{BKT}$/$t_{1}$ is plotted as a function of the pairing strength for both cases. In the strong-coupling regime, there is a significant amplification of $T_{BKT}$ in the two-band system compared to the single-band system. In both cases, the chemical potential is fixed around the vHs of the quasi-flat band. In the single-band case, the ratio between the mean-field pairing temperature $T_{c}$, that can be measured by ARPES or by the deviation from the normal-state resistivity lowering the temperature, and the $T_{BKT}$ cannot be lower than 10, indicating a strong suppression of $T_{BKT}$ and a very large pseudogap regime in temperature for single-band systems. Remarkably, we demonstrate that in the case of two bands, one deep and the other quasi-flat with a vHs, the $T_{BKT}$ is enhanced by an order of magnitude and the ratio $T_{c}$/$T_{BKT}$ can lower substantially toward values of the order 2-5.
Interestingly, FeSe superconducting monolayers show a BKT transition around 24 K \cite{zhang} and a mean-field temperature for onset of pairing and pseudogap opening, as detected by resistivity and ARPES, of the order of 55-65 K. Hence, the experimental ratio $T_{c}$/$T_{BKT}$ ranges from 2 to 3, a number that cannot be obtained with a single-band model system, but instead a two-band system such as the one considered in this work can allow ratios in the same range observed experimentally. 

\section{Conclusions}
In this work, we have explored the influence of electronic band configurations and their properties, such as van Hove singularities, and of the pairing regimes on the suppression of both superfluid stiffness and $T_{BKT}$ temperature. We have addressed the challenge of superfluid stiffness suppression by utilizing multi-band and multi-condensate effects, thereby enabling the stabilization of high-$T_{c}$ 2D superconductivity. In the single-band case, as the band becomes flatter, the effective mass of electrons increases, leading to an increasingly pronounced effect of superfluid stiffness suppression. This effect is even larger for 2D tight-binding electrons due to the presence of the van Hove singularity in the density of states. Interestingly, while the van Hove singularity is known to increase the mean-field superconducting critical temperature, it also leads to substantial suppression of the superfluid stiffness and hence of $T_{BKT}$. Additionally, we have observed an inverse relationship between superfluid stiffness and the superconducting gap in the strong-coupling regime. Consequently, the superconducting gap emerges as the primary determinant of superfluid stiffness suppression in the strong-coupling regime, which is generated in the vicinity of a van Hove singularity. Here, we have demonstrated a method for circumventing the suppression of the superfluid stiffness in quasi-flat band systems by providing a reservoir of phase stiffness with a deep band characterized by a weak pairing of its electrons, exploiting multi-band and multi-gap superconductivity. Specifically, we have analyzed a coupled quasi-flat and deep dispersing two-band 2D system by tuning the pair-exchange coupling between the band condensates. Our findings indicate that coupling with the deep band is able to increase the $T_{BKT}$ temperature by 10-100 times when compared to the single-band case and also significantly shrinks the pseudogap region between $T_{c}$ and $T_{BKT}$ arising because of amplitude, phase and vortex-antivortex fluctuations. We conclude that our proposed electronic configuration provides an effective means of screening the suppression of the superfluid stiffness and enhancing by order of magnitudes the $T_{BKT}$ temperature in the multiband coupled 2D systems that possess quasi-flat bands and strong van Hove singularities, coupled with deep bands having weak-pairing strengths.\\

\section*{Acknowledgements}
We are grateful to Tiago Saraiva (HSE), Giovanni Midei (University of Camerino), Giulia Venditti (University of Geneva) and Jonas Bekaert (University of Antwerp) for useful discussions and critical reading of the manuscript. This work has been supported by PNRR MUR project PE0000023-NQSTI and by Research Foundation-Flanders (FWO-Vl).

\end{document}